\documentclass[11pt]{article}
\usepackage[body={16.6cm,23cm}]{geometry}

\usepackage[hang,small,center]{caption}
\usepackage{amsmath,amssymb}
\usepackage{amsfonts}
\usepackage{mathrsfs}
\usepackage{dsfont}

\usepackage{epic}
\usepackage{eepic}
\usepackage{graphicx}

\usepackage{pgf,pgfarrows}
\usepackage{tikz}

\usepackage{cite}
\usepackage{enumerate}
\usepackage[curve]{xy}
\newcommand{\ds}{\displaystyle}

\renewcommand{\author}[1]{\large\rm #1\\ \bigskip}
\newcommand{\address}[1]{{\normalsize\it #1\\}\bigskip}
\renewcommand{\title}[1]{\bigskip\bigskip\Large\bf #1\bigskip\bigskip\\}

\newcommand{\Bigpsi}[3]{\phantom{\Psi}_2 \kern -.05em
\Psi_2\left(\genfrac{}{}{0pt}{}{#1}{#2}\biggl|#3\right)}

\newcommand{\bea}{\begin{eqnarray}}
\newcommand{\eea}{\end{eqnarray}}

\newcommand{\beq}{\begin{equation}}
\newcommand{\eeq}{\end{equation}}

\newcommand{\x}{{\boldsymbol{x}}}
\newcommand{\y}{{\boldsymbol{y}}}

\newcommand{\ii}{\mathsf{i}}

\newcommand{\ow}{\overline{\mathcal W}}
\newcommand{\w}{{\mathcal W}}
\newcommand{\s}{{\mathcal S}}

\newcommand{\cpar}{{\eta}}

\newcommand{\q}{{\mathsf q}}
\newcommand{\p}{{\mathsf p}}

\newcommand{\iW}{\mathcal{W}}
\newcommand{\iS}{\mathcal{S}}

\newcommand{\bb}{\mathsf{b}}

\def\EXP{\textrm{{\large e}}}

\newcommand{\url}[1]{}

\renewcommand{\textcolor}[1]{}

\newcounter{app}
\newcounter{sapp}[app]

\begin{document}

\vglue 2 cm
\begin{center}
\title{A new solution of the star-triangle relation}
\author{Andrew P.~Kels}
\address{Department of Theoretical Physics,
         Research School of Physics and Engineering,\\
    Australian National University, Canberra, ACT 0200, Australia.}

\vspace{2cm}

\begin{abstract}
We obtain a new solution to the star-triangle relation for an
Ising-type model with two kinds of spin variables at each lattice site,
taking continuous real values and arbitrary integer values, respectively.
The Boltzmann weights are manifestly real and positive. They are
expressed through the Euler gamma function and depend on sums and
differences of spins at the ends of the edge.

\end{abstract}

\end{center}

\newpage

The star-triangle relation is a distinguished form of the Yang-Baxter
equation for Ising-type models on two-dimensional lattices. In these
models the fluctuating variables or ``spins'' are assigned to lattice
sites, while two spins interact only if they are connected by an edge of
the lattice. Remarkably, many physically interesting models in
this class can be solved exactly, 
for instance, the 2-d Ising \cite{Bax82} and the chiral Potts
 \cite{AuY87,Baxter:1987eq}
models (see \cite{Bax02rip, BKS2} for a review of other known cases). The
star-triangle relation plays the role of the integrability condition
for these models. 

Recently, Bazhanov and Sergeev found a very important ``master''
solution \cite{BS10a} of the star-triangle relation, which contains all
previously known solutions of this relation as particular cases, and
provides many new interesting examples. The above master solution is
expressed in terms of the elliptic gamma function. It contains two
arbitrary free parameters $\p$ and $\q$, which play the role of
elliptic nomes. The spin variables take continuous real values
on the circle. Various interesting cases arise when the
parameters $\p$ and $\q$ approach some special values. These limits are
rather singular and their consideration requires a detailed analysis
on a case by case basis (for a comprehensive account see the forthcoming paper
\cite{BKS2}).  In the present paper we consider one particular
remarkable limit of the master solution, which leads to a model with
two types of 
spin variables at each lattice site.
These variables take continuous real values and
discrete integer values, respectively. Mathematically the master
solution of \cite{BS10a} is related to the elliptic beta integral
\cite{Spiridonov-beta}.    
Here we consider first a degeneration \cite{Spiridonov-statmech}
of this solution 
related to the ``hyperbolic beta integral'' \cite{Rains-limits, Stok04}.    
Then we consider the so-called ``strong coupling regime'', where new
integer-valued spin variables dynamically arise at each lattice site. 
Our calculations generalise the results of \cite{BSFV1}. 

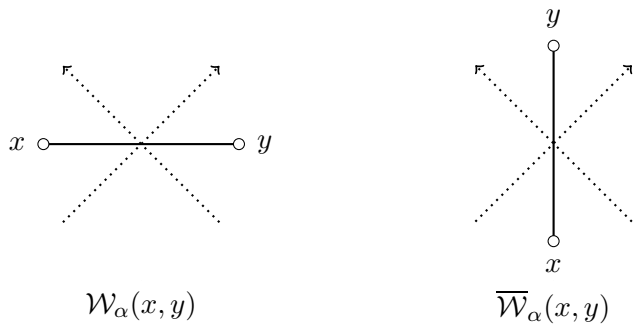
\begin{figure}[hbt]
\centering
\begin{tikzpicture}[scale=2.6]

\draw[-,thick] (-0.5,2)--(0.5,2);
\draw[->,thick,dotted] (0.4,1.6)--(-0.4,2.4);
\draw[->,thick,dotted] (-0.4,1.6)--(0.4,2.4);
\filldraw[fill=white,draw=black] (-0.5,2) circle (0.8pt)
node[left=3pt]{\color{black} $x$};
\filldraw[fill=white,draw=black] (0.5,2) circle (0.8pt)
node[right=3pt]{\color{black} $y$};

\fill (0,1.3) circle(0.01pt)
node[below=0.05pt]{\color{black} $\w_{\alpha}(x,y)$};

\begin{scope}[xshift=60pt,yshift=57pt]
\draw[-,thick] (0,-0.5)--(0,0.5);
\draw[->,thick,dotted] (-0.4,-0.4)--(0.4,0.4);
\draw[->,thick,dotted] (0.4,-0.4)--(-0.4,0.4);
\filldraw[fill=white,draw=black] (0,-0.5) circle (0.8pt)
node[below=3pt]{\color{black} $x$};
\filldraw[fill=white,draw=black] (0,0.5) circle (0.8pt)
node[above=3pt]{\color{black} $y$};

\fill (0,-0.7) circle(0.01pt)
node[below=0.05pt]{\color{black} $\ow_{\alpha}(x,y)$};
\end{scope}
\end{tikzpicture}
\caption{Edges of the first (left) and second types and their Boltzmann weights.}
\label{2boltzmannweights}
\end{figure}

Consider the square lattice of $N$ sites and assign spin variables
$x_j$, $j=1,2,\ldots,N$, taking some continuous set of values, to all
lattice sites.  
Two spins interact only if they are connected by an edge. Let 
$\w_\alpha(x,y)$, $\ow_\alpha(x,y)$ denote Boltzmann weights for
horizontal and vertical edges, where $x$ and $y$ are spins at the end
of the edge, as shown in Figure~\ref{2boltzmannweights}. 
 We assume that the edge weights depend on the (additive) spectral
variable $\alpha$ and are related to each other by the crossing symmetry
$\ow_\alpha(x,y)=\w_{\eta-\alpha}(x,y)$, where $\eta$ is the model
dependent crossing parameter. Moreover we assume that the weights are
reflection symmetric $\w_\alpha(x,y)=\w_\alpha(y,x)$. It is also
convenient to introduce single-spin weights $\s(x_j)$, which are
independent of the spectral variable $\alpha$.  
The partition function is defined as 
\begin{align}
\label{z-main}
{\cal Z}=\mathop{\idotsint}\
\prod_{(ij)}\w_{\alpha}(x_i,x_j)\
\prod_{(kl)}\w_{\cpar-\alpha}(x_k,x_l)\ \prod_{m}
\s(x_m)\,d x_m\;,
\end{align}
where the first product is taken over all horizontal edges $(ij)$, the second
over all vertical edges $(kl)$ and the third over all internal sites of the
lattice.
We will implicitly assume fixed boundary conditions. Obviously the
single spin weights $\s(x_j)$ can be included into the definition of
the edge weights, but we prefer not to do so. The model is integrable
if the weights satisfy the star-triangle equation of the form 
\begin{align}
\label{msstr}
\begin{array}{l}
\ds\int
dx_0\,\iS(x_0)\iW_{\eta-\alpha_1}(x_1,x_0)\iW_{\eta-\alpha_2}(x_2,x_0)\iW_{\eta-\alpha_3}(x_3,x_0)\\[.3cm]
\phantom{MMMMMMMMM}\ds=
{\cal
  R}(\alpha_1,\alpha_2,\alpha_3)\,\iW_{\alpha_1}(x_2,x_3)\iW_{\alpha_2}(x_1,x_3)\iW_{\alpha_3}(x_2,x_1)\;,
\end{array}
\end{align}
where the spectral parameters $\alpha_1$, $\alpha_2$, $\alpha_3$
satisfy the relation $\alpha_1+\alpha_2+\alpha_3=\eta$ and the factor $  
{\cal   R}(\alpha_1,\alpha_2,\alpha_3)$ is independent of the spins
$x_1,x_2,x_3$.

Let us now define the Bazhanov-Sergeev master solution \cite{BS10a}.
Introduce the elliptic nomes $\p,\q$, and the crossing parameter $\eta$
\begin{align}
\q=\EXP^{\ii\pi\tau}\;,\quad\p=\EXP^{\ii\pi\sigma}\;,\quad |\,\p\,|\;,\;|\,\q\,|\,<\,1\;,
\quad\EXP^{-2\eta}=\p\q\;,\quad\eta=-\frac{\ii\pi}{2}(\tau+\sigma)\;.
\end{align}
The elliptic gamma function \cite{Ruijsenaars-elliptic,FelVar00} is defined as
\begin{align}
\label{EGF}
\Phi(z)=\prod^\infty_{j,k=0}\frac{1-\EXP^{2\ii z}\q^{2j+1}\p^{2k+1}}{1-\EXP^{-2\ii z}\q^{2j+1}\p^{2k+1}}
=\exp\left\{\sum_{n\neq0}\frac{\EXP^{-2\ii n x}}{n(\p^n-\p^{-n})(\q^n-\q^{-n})}\right\}\;,
\end{align}
where the exponential form is valid in the strip
$|\,\mbox{Im}\,z\,|\,<\mbox{Re}\,\eta$. 
The Boltzmann weights are defined by
\begin{align}
\label{msdef}
\iS(x)=\ds\frac{\EXP^{\eta/2}}{2\pi}
\vartheta_1(2x\,|\,\p)\vartheta_1(2x\,|\,\q)\;,\quad\w_\alpha(x,y)
=
\ds\kappa(\alpha)^{-1}\frac{\Phi(x+y+\ii\alpha)
\Phi(x-y+\ii\alpha)}{\Phi(x+y-\ii\alpha)\Phi(x-y-\ii\alpha)}\;, 
\end{align}
where $\vartheta_1(z\,|\,\q)$ is the standard Jacobi theta function
\cite{WW}. Note, that the weights are periodic 
\begin{align}
\w_\alpha(x+\pi,y)=\w_\alpha(x,y+\pi)=
\w_\alpha(x,y),\qquad \iS(x+\pi)=\iS(x).
\end{align}
 Correspondingly,
the spin variables in \eqref{z-main} take values in the
interval\  $0\leq x_m<\pi$. 
The normalisation factor $\kappa(\alpha)$ in \eqref{msdef} reads
\begin{align}
\label{kapnorm}
\log\kappa(\alpha)=\sum_{n\neq0}\frac{\EXP^{4\alpha n}}{n(\p^n-\p^{-n})(\q^n-\q^{-n})(\p^n\q^n+\p^{-n}\q^{-n})}\;.
\end{align}
The above Boltzmann weights satisfy the star-triangle relation \eqref{msstr} 
where the integration is taken over the segment $0\leq x_0<\pi$. 
Below we assume that the spectral parameter $\alpha$ lies in the
domain $0<\alpha<\eta$. In this case the weights \eqref{msdef} are
real and positive.   

With
the normalisation \eqref{kapnorm} the factor $\mathcal{R}$ in
\eqref{msstr} is equal to one.  For the same normalisation, the 
weights \eqref{msdef} also satisfy the following inversion
relations 
\begin{align}
\label{msinv}
\begin{array}{rcl}
\ds\w_\alpha(x,y)\w_{-\alpha}(x,y)&=&1\;, \\[0.3cm]
\ds\int^{\pi}_0dz\,\s(z)\w_{\eta-\alpha}(x,z)\w_{\eta+\alpha}(z,y)&=&\ds\frac{1}{2\,\s(x)}(\delta(x+y)+\delta(x-y))\;.
\end{array}
\end{align}
These relations allow one to show \cite{BS10a} that in the thermodynamic limit,
when the number of lattice sites goes to infinity $N\to\infty$, the
bulk free energy of the system vanishes,
\begin{align}
 \lim_{N\to \infty} N^{-1} \log{\cal  Z} = 0\,.\label{fzero}
\end{align}

As a mathematical identity the star-triangle relation \eqref{msstr} is equivalent 
to the elliptic beta integral \cite{Spiridonov-beta}, which plays a fundamental 
role in the theory of elliptic hypergeometric functions.  Its significance for 
solvable models of statistical mechanics was discovered in \cite{BS10a}.

Now consider the hyperbolic limit \cite{Spiridonov-statmech} of \eqref{msdef}.
Introduce the modular parameter $\bb$ to take values in one of the
following two regimes 
\begin{align}
(i)\;\bb>0\;,\qquad(ii)\;|\,\bb\,|=1\;,\quad\mbox{Im}\,(\bb^2)>0\;.
\end{align}
The non-compact quantum dilogarithm is defined as the integral
\begin{align}
\label{ncqddef}
\varphi(z)=\exp\left\{\frac{1}{4}\int_{\mathbb{R}+\ii0}\frac{\EXP^{-2\ii yz}dy}{y\sinh(y\bb)\sinh(y/\bb)}\right\}\;,
\quad |\,\mbox{Im}\,(z)\,|\,<\,\mbox{Re}\,(\eta)\;,
\end{align}
where the singularity at the origin is taken below the contour.  
Substituting 
\begin{align}
\label{varchange}
\p=\EXP^{-b\epsilon}\;,\quad\q=\EXP^{-b^{-1}\epsilon}\;,\quad x_i
=x_i\epsilon\;,\quad\alpha_i
=\alpha_i\epsilon\;,\quad i=1,2,3\;,
\end{align}
into \eqref{msdef} 
and taking the limit $\epsilon\rightarrow0$, one obtains
\cite{Spiridonov-statmech}, 
\begin{align}
\label{genfadvoldef}
\iS(x)=\ds2\sinh(2\pi x_0\bb)\sinh(2\pi
x_0/\bb)\;,\quad\iW_\alpha(x,y)=\ds\kappa(\alpha)^{-1}\EXP^{4\pi\alpha
  x}\frac{\varphi(x+y+\ii\alpha)}{\varphi(x+y-\ii\alpha)}\frac{\varphi(x-y+\ii\alpha)}{\varphi(x-y-\ii\alpha)}\;.
\end{align}
These weights satisfy the star-triangle relation \eqref{msstr}, where 
the integration taken over the whole real line and the crossing parameter
is given by $\eta=(\bb+\bb^{-1})/2$.  
The normalisation factor $\kappa(\alpha)$ in 
\eqref{genfadvoldef} reads 
\begin{align}
\label{kaphyp}
\log\kappa(\alpha)=\frac{1}{8}\int_{\mathbb{R}+\ii0}
\frac{\EXP^{4yz}dy}{y\sinh(y\bb)\sinh(y/\bb)\cosh(2y\eta)}
+\pi\ii\alpha^2-\frac{\pi\ii}{3}\eta^2+\frac{\pi\ii}{24}\;.
\end{align}
The weights \eqref{genfadvoldef} define a physical model of
statistical mechanics, rather similar to the Faddeev-Volkov model
\cite{FV95,BSFV2}. The spins now take arbitrary values on the real line
$x_j\in{\mathbb R}$. The above limiting procedure does not affect the
bulk free energy of the system, so that the relation \eqref{fzero}
continues to hold for the new model. 

Let us now consider a further specialisation of the above 
generalised Faddeev-Volkov model \eqref{genfadvoldef}, 
when the parameter $\bb\rightarrow\ii$, or equivalently, $\eta\rightarrow0$.
In this limit the weights $\iW_\alpha(x,y)$ 
develop a series of delta function like peaks for integer values 
of the sum $x+y\in\mathbb{Z}$ or differences $x-y\in\mathbb{Z}$
of the arguments $x,y$.  This limit was previously 
investigated for the Faddeev-Volkov model \cite{BSFV1} and a similar
phenomenon was observed.  A correct limiting procedure capturing a
fine structure of these peaks requires a
redefinition of the spin variables. 
The asymptotics of the non-compact quantum dilogarithm $\varphi(z)$
and the function $\kappa(\alpha)$ read
\begin{align}
\label{varchange2}
\varphi(n+x\eta)=\EXP^{-\ii\pi/12}(4\pi\eta)^{\,\ii
  x}\,\frac{\Gamma(\frac{1-n+\ii x}{2})}{\Gamma(\frac{1-n-\ii
    x}{2})}\;,\quad\kappa(\beta\eta)=(8\pi\eta)^{\,\ii
  \beta}\,\frac{\Gamma(\frac{1+\ii \beta}{2})}{\Gamma(\frac{1-\ii
    \beta}{2})}\;, 
\end{align}
where $\eta\to0$,\ \ $n\in\mathbb{Z}$, \ \ $x\ll\eta^{-1}$ and
$0<\beta<1$.  Using this in \eqref{genfadvoldef} one obtains 
the following $\eta\to0$ asymptotics,\footnote{Here we use the compact
  notation, 
  where the $\pm$
  symbol in the argument of the gamma function indicates the product
  of gamma functions with both signs are taken as
  e.g. $\Gamma(a\pm b)\equiv\Gamma(a+b)\Gamma(a-b)$} 
\begin{align}
\label{smsg}
\begin{array}{rcl}
\iS(n+x\eta)&\!\!\!\!\to\!\!\!\!\!&\ds8\pi^2(x^2+n^2)\,\eta^2\;, \\[0.3cm]
\iW_{\beta\eta}(m+x\eta,n+y\eta)&\!\!\!\!\to\!\!\!\!\!&
\ds\frac{2^{-5\beta}}{(\pi\eta)^{-3\beta}}\,\frac{\Gamma(\frac{1+\beta}{2})}{\Gamma(\frac{1-\beta}{2})}\,
\frac{\Gamma(\frac{1-\beta}{2}\pm\frac{\ii(x+y)-(m+n)}{2})\,\Gamma(\frac{1-\beta}{2}\pm\frac{\ii(x-y)-(m-n)}{2})}
{\Gamma(\frac{1+\beta}{2}\pm\frac{\ii(x+y)+(m+n)}{2})\,\Gamma(\frac{1+\beta}{2}\pm\frac{\ii(x-y)+(m-n)}{2})}\;,
\end{array}
\end{align}
where $m,n\in\mathbb{Z}$ and $x,y\in\mathbb{R}$. 
Substituting this into \eqref{msstr} and cancelling out
singular factors from both sides, one obtains the following
star-triangle relation
\begin{align}
\label{strmsg}
\begin{array}{r}
\ds\sum_{n_0\in\mathbb{Z}}\;\int^{\infty}_{-\infty}dx_0\,
\iS(x_0,n_0)\iW_{1-\beta_1}(x_1,n_1\,|\,x_0,n_0)\iW_{1-\beta_2}(x_2,n_2\,|\,x_0,n_0)
\iW_{1-\beta_3}(x_3,n_3\,|\,x_0,n_0) \\\ds=\iW_{\beta_1}(x_2,n_2\,|\,x_3,n_3)
\iW_{\beta_2}(x_1,n_1\,|\,x_3,n_3)\iW_{\beta_3}(x_2,n_2\,|\,x_1,n_1)\;,
\end{array}
\end{align}
where $\beta_1+\beta_2+\beta_3=1$, and the Boltzmann weights are given by
\begin{align}
\label{2spinwts}
\iS(x,n)=\ds\frac{1}{2\pi}(x^2+n^2)\;,\quad\iW_{\beta}(x,n\,|\,y,m)
=\ds\frac{\Gamma(\frac{1+\beta}{2})}{\Gamma(\frac{1-\beta}{2})}\,
\frac{\Gamma(\frac{1-\beta}{2}\pm\frac{\ii(x+y)-(m+n)}{2})\,\Gamma(\frac{1-\beta}{2}\pm\frac{\ii(x-y)-(m-n)}{2})}
{\Gamma(\frac{1+\beta}{2}\pm\frac{\ii(x+y)+(m+n)}{2})\,\Gamma(\frac{1+\beta}{2}\pm\frac{\ii(x-y)+(m-n)}{2})}\;.
\end{align}
These weights give a new solution of the
star-triangle relation for an Ising type model, 
where sites of the lattice are assigned ``dual spin'' variables $\x$ 
\begin{align}
\label{2spinvars}
\x=(x,n)\;,\qquad x\in\mathbb{R}\;,\quad n\in\mathbb{Z}\;,
\end{align}
taking arbitrary real and integer values, respectively.
The Boltzmann weights \eqref{2spinwts}
satisfy the spin reflection symmetry $\x\leftrightarrow\y$, for two spins 
$\x,\y$ connected by an edge as in Figure~\ref{2boltzmannweights}.  
The bulk free energy \eqref{fzero} of the
corresponding 2-d lattice model remains unchanged.

The star-triangle relation \eqref{strmsg} is derived in the strong-coupling limit
 of the generalised Faddeev-Volkov model \eqref{genfadvoldef},
the latter model arising as a hyperbolic degeneration of the master solution
\eqref{msdef}.  From either of the models \eqref{msdef}, \eqref{genfadvoldef}, new 
solutions of the star-triangle relation can be found as different singular
limits of the parameters $\p,\q$, and $\bb$ respectively.  In each case, the spins 
and spectral parameters should be redefined, as was done here in \eqref{varchange} and 
\eqref{smsg}, in order to obtain convergent limits.

\section*{Acknowledgments}
I thank Vladimir Bazhanov for suggesting the problem and useful advice.

\bibliography{2spin}
\bibliographystyle{vvb-bibstyle}

\end{document}